# Corporate Transparency and the Disposition Effect[1]

**Siliu Chen, Fei Ren** *

School of Business, East China University of Science and Technology, Shanghai, China

*** Correspondence:**
Fei Ren
fren@ecust.edu.cn

## Abstract

The disposition effect describes investors' irrational behavior of selling profitable assets too soon while holding onto losing assets for too long. This study examines the impact of transparency at the firm level on the disposition effect of individual investors who hold that company's stock. Our results show that an increase in corporate transparency significantly reduces the disposition effect. Further analysis reveals that for companies with greater transparency, when the held stock is profitable, investors' confidence in holding it increases, leading to a reduced bias toward selling profitable stocks. When the stock is held at a loss, investors' confidence in holding it weakens, but they often perceive the loss as temporary and maintain confidence in the company's long-term prospects, thus exacerbating the bias toward holding losing stocks. The effect of increased transparency on the selling behavior of profitable stocks is greater than its effect on the selling behavior of losing stocks. Overall, an increase in corporate transparency significantly reduces the disposition effect.

## 1 Introduction

The disposition effect refers to a persistent phenomenon in which investors tend to hold their losers too long and sell their winners too soon. This behavioral anomaly, which contradicts the rational person assumption and expected utility theory of traditional finance, was first proposed by Shefrin and Statman (1985). Subsequent studies have shown the disposition effect is widespread and has been observed for stocks (Wu et al., 2020, Odean, 1998, Lu et al., 2022), funds (Wu et al., 2016, Cici, 2012), and futures (Locke and Mann, 2005, Cheng et al., 2024, Huang and Tsai, 2025). Numerous studies examine the relationship between investor characteristics and the disposition effect. Experienced investors tend to make more accurate predictions about the direction of the stock market and exhibit a lower disposition effect (Xiao et al., 2018, Pereira da Silva and Mendes, 2021). Investors with higher levels of education also tend to show a lower disposition effect (Pereira da Silva and Mendes, 2021). Investors in first-tier cities in China exhibit a stronger disposition effect compared to those in smaller cities (Wu et al., 2016). Compared to male investors, female investors generally have less financial knowledge (Jiang et al., 2020), higher risk aversion (Mohammadi and Shafi, 2018), and are more prone to regret (Cao et al., 2021), resulting in a stronger disposition effect (Wu et al., 2016, Xiao et al., 2018, Cao et al., 2021). Ben-David and Hirshleifer (2012) examine the disposition effect and show the probability of selling follows an asymmetric V-shaped pattern – it increases with both gains and losses but rises more rapidly for profitable stocks. The probability of selling profitable stocks was higher than the probability of selling losing stocks, confirming the disposition effect.

[1] Published in Frontiers in Psychology. DOI: https://doi.org/10.3389/fpsyg.2025.1626829

Given that individual investors exhibit this disposition effect, it is useful to seek ways to reduce it. Improvements in the information environment help to reduce information asymmetries, allowing investors to make better investment decisions, thereby improving market efficiency (Chen and Wu, 2022). Chen and Ren (2025) find that information on social media platforms significantly improves the information environment for investors, helping to reduce the disposition effect. Yin and Zhu (2025) show that information obtained directly from companies is more important to investors than information exchanged on social media platforms. Liu et al. (2023) find that greater transparency in companies provides investors with more comprehensive information, which helps control risks and strengthen investor confidence, ultimately improving firm value. Based on this, we hypothesize that when a company's information transparency is higher, the information environment for individual investors improves, asymmetry of information decreases, and investors can make more rational judgments about the stock, thereby reducing the disposition effect for that stock. However, to our knowledge no studies examine the impact of corporate transparency on the disposition effect exhibited by individual investors holding that stock. We address this gap in the literature by examining, from the perspective of corporate transparency, whether individual investors' disposition effects vary across stocks of different companies, and how these differences arise.

We use real trading data from individual investors on the popular Chinese social investment platform Xueqiu to investigate the impact of corporate transparency on individual investors' disposition effect and reveals its underlying mechanism. Our results show that increased corporate transparency can significantly reduce the disposition effect for individual investors who hold that company's stock. Further analysis reveals that for companies with greater transparency, when the held stock is profitable, investors' confidence in holding the stock increases, leading to a reduced bias toward selling profitable stocks. When the held stock is held at a loss, investors' confidence in holding the stock weakens, but they often perceive the loss as temporary and maintain confidence in the company's long-term prospects, thus exacerbating the bias toward holding losing stocks. The effect of increased transparency on the selling behavior of profitable stocks is greater than its effect on the selling behavior of losing stocks, thus significantly reducing the disposition effect.

The contributions of this study are as follows: First, this study is the first to examine the impact of corporate transparency on the disposition effect for individual investors, finding that an increase in transparency significantly reduces the disposition effect for the company's stock. This provides a new perspective on the disposition effect and enriches the related theoretical research. Second, we clarify the mechanism through which corporate transparency influences the disposition effect for a stock. We found that it is the asymmetric confidence shown towards profitable and losing stocks that causes the probability of selling profitable stocks to be much lower than the probability of selling losing stocks, thereby reducing the disposition effect.

## 2 Literature Review and Research Hypothesis

### 2.1 Disposition Effect

Shefrin and Statman (1985) are the first to identify and propose the concept of the disposition effect, which describes investors' tendency to hold onto losing assets too long and sell winning assets too soon. Odean (1998) studies data from 10,000 accounts provided by a nationwide discount brokerage firm in the U.S. from 1987 to 1993 and proposes the classic disposition effect measure, PGR-PLR.

He finds the proportion of profitable stocks sold is higher than that of losing stocks, demonstrating the existence of the disposition effect in the U.S. stock market. Xiao et al. (2018) analyze transaction data from 30,512 accounts provided by a brokerage firm in China, using the classic PGR-PLR measure and the Cox proportional hazard model to demonstrate the existence of the disposition effect in the Chinese securities margin trading. They investigate the factors influencing the disposition effect through a survival analysis and find that gender, age, and investment level all affect an investor's tendency to display the disposition effect. Specifically, female investors exhibit a stronger disposition effect compared to male investors, and middle-aged investors have the strongest disposition effect, followed by young investors, while elderly investors display the lowest disposition effect. Investors with higher investment sophistication tend to exhibit a lower disposition effect. Danbolt et al. (2022) analyze variations in trading behaviors within and between private and publicly visible portfolios using proprietary data from a European fintech social trading platform. They find the disposition effect diminishes by roughly 35% when trades and holdings are public and conclude that the level of transparency regarding trading activities and portfolio holdings, as well as the way financial information is shown, can influence trading decisions. Bachmann (2024) divide risk-takers into two groups based on whether they have the autonomy to choose how many shares of risky assets to hold. He finds that after experiencing losses, risk-takers with that autonomy held more optimistic beliefs about the prospects for their investments than risk-takers without a choice, despite having identical risk exposures. Risk-takers with a choice held their losing positions longer, while those without a choice sold their losing assets more quickly. This led to the recognition of the role of belief-updating in the disposition effect. An et al. (2024) explore the relationship between portfolio returns and the disposition effect, finding the disposition effect for a stock significantly weakens if the overall portfolio shows a gain, but is large when the portfolio is at a loss. In addition, studies that examine the relationship between the disposition effect and returns find the stronger the disposition effect, the greater the losses (Cheng et al., 2024, Fan and Neupane, 2024, Huang and Tsai, 2025).

## 2.2 Corporate Transparency

Existing studies examine the impact of corporate transparency on stock market pricing efficiency. Xin et al. (2014) investigate the effect of corporate transparency on stock price volatility, finding that greater transparency reduces stock price volatility. They argue that a more transparent information environment lowers information risk and improves valuation accuracy. Firth et al. (2015) examine how corporate transparency helps to explain the sensitivity of stock prices to investor sentiment. They find that the lower the level of corporate transparency, the greater the influence of sentiment on stock prices. They also show that greater corporate transparency can mitigate subjective influences on stock pricing, thus improving market efficiency. Xu et al. (2023) show that corporate transparency is negatively correlated with stock price synchronicity. Similarly, Xiang and Lu (2020) find the transparency of Chinese A-share companies is significantly negatively associated with post-earnings announcement drift. In addition, Liu et al. (2023) find corporate transparency is significantly positively related to firm value. Specifically, greater transparency provides more comprehensive information to investors, helping to more effectively control risks and strengthening investors' confidence, thereby improving the firm value.

Improvements in the information environment help to reduce information asymmetry, allowing investors to make better investment decisions, increasing market efficiency (Chen and Wu, 2022). Compared to the information exchange between investors on social platforms, information obtained directly from companies is more important for investors (Yin and Zhu, 2025). Greater transparency in companies provides investors with more comprehensive information, which helps control risks and strengthen investor confidence, ultimately improving firm value (Liu et al., 2023). Building on this review of the literature, we hypothesize that when company-level information transparency is high, the information environment for individual investors improves, asymmetry of information decreases, and investors can make more rational judgments about the stock, thereby reducing the disposition effect for that stock. Therefore, we propose the following hypothesis:

Hypothesis: An increase in corporate transparency can reduce the disposition effect of individual investors who hold the company's stock.

# 3 Research Methods

## 3.1 Sample and Data

We use actual trading data from Xueqiu (https://xueqiu.com/) to calculate the disposition effect for individual investors. Xueqiu was established on November 11, 2011, and has become one of the most active social investment platforms in China. On Xueqiu, users can choose to create real trading portfolios based on actual trades that can be viewed by other users as a reference for their own portfolio strategies. We used a Python-based web crawler to collect transaction data from real trading portfolios on Xueqiu, covering 11,661 portfolios for the period from June 27, 2016, through May 31, 2023. Our dataset includes portfolio ID, transaction time, traded stocks, transaction price, stock holdings before the transaction, and stock holdings after the transaction. After obtaining the raw data, we performed the following data cleaning steps: (1) As Xueqiu does not provide complete transaction records for real trading users but instead offers only the latest 200 transactions per user, we excluded transactions where the purchase price for a stock could not be obtained; (2) Transactions involving new stock subscriptions and their subsequent sales were excluded; (3) Only transactions involving A-shares were included; (4) Transactions with obvious errors were excluded; (5) Portfolios with multiple transactions containing obvious errors were excluded; (6) As changes in stock codes are typically associated with significant company changes, trades involving such changes are excluded. After these cleaning steps, our final dataset contained 1,115,839 transaction records from 11,323 real trading portfolios.

Next, we organized the trading data for each trading portfolio. For individual stock trading days (i.e., non-suspension days), if the investor made a trade we record the profit and loss status of the stock before the trade and the trade direction for each transaction. If no trades were made, we mark the profit and loss status at the close and record the day as a holding status. The profit and loss status before a trade is determined by comparing the current trading price with the purchase price (in the case of multiple purchases, we use the weighted average purchase price). If the trading price is higher than the purchase price, it is classified as a profitable sale; if the trading price is lower than the purchase price, it is considered a loss; if the trading price is equal to the purchase price, it is marked as a zero return status. The profit and loss status at the close is determined by comparing the day's closing price with the purchase price. When there are multiple purchases of the same stock, we use the weighted average purchase as the price purchase. The purchase price is adjusted for any stock dividends.

The average number of trades per day for a single investor on the same stock is 1.1695, with a median of 1.0 and a standard deviation of 0.5672. After incorporating holding period data into the transaction-level data, we obtain a total of 14,496,822 observations. Each observation corresponds to one of four states for a stock: opening a position, adding to a position, reducing a position (including closing a position), or holding a position. Since T+1 settlement is used in the Shanghai and Shenzhen stock markets, stocks bought on a given day can only be sold on the next trading day or any subsequent trading day, meaning they cannot be sold the day of purchase. Therefore, we exclude opening day data, which could not reflect investors' disposition effects, from the regression analysis. This left us with 14,112,512 observations.

The data used to construct a firm transparency index (as described below), along with the daily closing price, highest price, lowest price of each stock, and company financial data, are sourced from the CSMAR database. Historical dividend information and the closing prices of the CSI 300 Index are sourced from the RESSET database.

### 3.2 Model Specification and Variable Definition

We draw on the methods in Chang et al. (2016), Frydman and Wang (2020), and Lu et al. (2022) to examine whether there is a disposition effect for individual investors in China using the following regression:

$$\boldsymbol{Sell_{ijt} = \beta_0 + \beta_1 Gain_{ijt} + Controls_{ijt} + \varepsilon_{ijt}}$$

In this model, since each transaction may reflect different psychological factors of the investor, we perform the regression based on the dimensions of investor *i*, stock *j*, and transaction *t*. For trading days with no transactions during the holding period, we include the holding status in the regression. *Sell* and *Gain* are dummy variables, with *Sell* set to 1 when a stock is sold, and 0 otherwise. *Gain* is set to 1 when the return for the position is positive, and 0 otherwise. The control variables (*Controls*) include the number of holding days, purchase price, the volatility of stock returns over the past 250 trading days, the market return (the return on the CSI 300) and its volatility over the past 20 trading days, the company's total market value, debt-to-asset ratio, book-to-market ratio, and return on assets at the end of the previous year. $\varepsilon$ is the random disturbance term. As the dependent variable is binary, we primarily use a Probit model for the regression analysis. Our main focus is on the coefficient $\beta_1$ of the independent variable *Gain*, which represents the increase in the probability of selling the stock when the investor holds the stock at a gain, compared to when it is held at a loss. If $\beta_1$ is significantly greater than 0, it indicates the existence of disposition effect, and the magnitude of $\beta_1$ reflects the size of this effect. Definitions for the variables are shown in Table 1.

Based on the above regression, we examine the impact of firm-level factors on individual investors' disposition effects from the perspective of corporate transparency. To avoid the limitations of using a single measurement indicator (Xin et al., 2014, Lang et al., 2012), we construct a corporate transparency index based on five dimensions: earnings quality, information disclosure evaluation, the number of analysts covering the company, analysts' earnings forecast accuracy, and audit firm quality. Next, we aggregate the annual corporate transparency index values for the stocks in our sample, calculate the transparency quartiles for each year, and categorize stocks into groups based on their firm's level of transparency. We run the regression on each group and the heterogeneity of results across the different groups reflects the variation in disposition effects of individual investors on stocks with varying levels of corporate transparency.

The process for constructing the corporate transparency index is as follows: The first transparency indicator is earnings quality, measured as the absolute value of discretionary accruals, calculated using the adjusted Jones model. To make this measure consistent with our other transparency indicators, we multiply the absolute value of discretionary accruals by -1 to define the earnings quality (*EQ*) indicator. The larger the *EQ*, the less room for earnings management and the higher the earnings quality.

The second transparency indicator is the evaluation score of a companies' information disclosure by the Shanghai and Shenzhen Stock Exchanges. The results are classified into four levels (A, B, C, D) based on the quality of disclosures, ranging from high to low, with corresponding scores of 4, 3, 2, and 1, respectively. This score is the information disclosure evaluation indicator (*Score*), where a higher score indicates higher quality disclosures.

The third transparency indicator is the number of analysts covering the company, specifically the number of analysts who make annual earnings forecasts for the company in a given year (*AF*). A higher number of analysts enriches a company's information environment. Therefore, more analysts who cover the company, the greater the transparency.

The fourth transparency indicator is the accuracy of analysts' earnings forecasts. To calculate this measure we first find the median value of the earnings per share (EPS) forecasts made by the analysts who follow the company for a given year. We then subtract the actual EPS, divide the result by the stock price from the previous year, take the absolute value, and multiply this by -1. The resulting value is the earnings forecast accuracy indicator (*FA*). The larger the value, the more accurate the analyst's earnings forecast, and the higher the transparency.

The fifth indicator is the quality of the audit firm. Existing research suggests the audit quality of the Big Four accounting firms is higher. When a company hires one of the Big Four to audit its annual report, it signals that the company is more likely to provide fair and accurate accounting and internal control information, which may indicate greater transparency. Therefore, if a listed company hires an international Big Four accounting firm to audit its annual report, the variable *Big4* is coded as 1; otherwise, it is coded as 0.

Based on the five transparency indicators described above, we follow the approach in Lang et al. (2012) and Xin et al. (2014) to assign the transparency values for the sample companies into percentiles. The average of the percentiles for the five indicators is then used as the firm's transparency indicator, *Transpr*. The higher the value of *Transpr*, the greater the firm's transparency.

## 4 Empirical Results

### 4.1 Descriptive Statistics

Table 2 presents descriptive statistics for the variables, reflecting the overall status of individual investors' stocks positions during trading and holding periods. The mean value of the dependent variable *Sell* is 0.0337, the mean value of the independent variable *Gain* is 0.3880, and the mean value of *Transpr* is 62.8567. Table 3 presents a correlation analysis of variables in the baseline regression, showing there is no strong correlation between any of the variables.

### 4.2 Baseline Regression

We use Probit and Logit models, which are suitable for binary dependent variables, to examine whether there is a noticeable disposition effect for individual investors. The results are presented in Table 4. Probit and Logit regression models only provide limited information, such as significance and parameter signs; they cannot directly show the impact of independent variables on the dependent variable. Therefore, Table 4 presents the average marginal effects (AME) of each independent variable on the dependent variable using both the Probit and Logit models, shown in the columns labeled Probit_AME and Logit_AME, respectively. The regression results using the Probit model show that when individual investors' stock returns change from negative to positive, the probability of selling the stock increases significantly, by 0.0226. Similarly, the regression results from the Logit model indicate that when individual investors' stock returns change from negative to positive, the probability of selling the stock increases significantly by 0.0220. Therefore, we conclude the disposition effect for individual investors is significant, and the magnitude of this effect, as calculated by both the Probit and Logit models, is similar, indicating the robustness of the results. Furthermore, the regression results of the control variables as shown in Table 4 indicate that as the length of time a stock is held (*sqrt_TimeOwned*) increases, the probability of selling the stock declines significantly while a higher buy price (*log_BuyPrice*) significantly increases the probability a stock will be sold. This suggests that as the number of days since a purchase was made increases, individual investors are more likely to hold the stock long-term, while the higher the buy price of the stock, the more likely they are to sell it.

After confirming the existence of a significant disposition effect for individual investors, we aggregate the transparency data of the companies corresponding to the stocks in the sample on an annual basis, and calculate annual transparency quartiles. The trading data is then grouped according to the level of corporate transparency corresponding to each stock and we conduct a regression analysis for each grouping. Table 5 presents the average marginal effects of the independent variables based on the Probit model when corporate transparency is classified as low, relatively low, relatively high, and high. The average marginal effect of *Gain* decreases sequentially across these groups, suggesting that as the return on a held stock changes from negative to positive, the incremental probability of selling the stock gradually declines. In other words, as corporate transparency increases, individual investors' disposition effect tends to decrease. These results validate our research hypothesis.

### 4.3 Robustness Tests

In the baseline analysis we conducted grouped regressions based on corporate transparency quartiles to examine the disposition effect for individual investors holding stocks in companies with different levels of transparency. Regression models that incorporate interaction terms are common, but may suffer from multicollinearity issues. As a robustness check, we construct a corporate transparency dummy variable, *Transpr_dv*, based on the median level of corporate transparency for each year. Companies above the median are assigned a value of 1, and those below the median are defined as 0. We add *Transpr_dv* and its interaction term with *Gain* into the regression equation. The VIF for *Gain* is 6.12 and the VIF for the interaction term between *Gain* and *Transpr_dv* is 6.93, while the VIF for all other variables is below 2.0. The regression results, presented in Table 6, show the coefficient and average marginal effect of the interaction term between *Gain* and *Transpr_dv* are significantly negative at the 1% level. This suggests that greater corporate transparency can reduce the disposition effect for investors holding those stocks, and thus our baseline results remain robust.

In addition, given that the baseline regression is based on four groups while the analysis above is based on two groups determined by the median level of corporate transparency, we include two-group regressions in the robustness check and calculate the inter-group differences. The results are shown in Table 6. The grouped regression results show the disposition effect is smaller in the high transparency group than in the low transparency group. Suest test results show the difference in the regression coefficients for *Gain* between the two groups is significant at the 1% level, confirming the robustness of the baseline regression results.

There is a potential for sample selection bias in grouped regressions, as individuals within different groups may differ. Experienced investors may tend to choose stocks with higher corporate transparency, and could exhibit a lower disposition effect compared to inexperienced investors (Xiao et al., 2018, Pereira da Silva and Mendes, 2021). Therefore, it is unclear whether the lower disposition effect in the high corporate transparency group is influenced by corporate transparency or investment experience. We use propensity score matching to address this sample selection bias issue across the different groups. We use the natural logarithms of the number of followers on the Xueqiu platform, the number of posts on the Xueqiu platform, and the number of stocks followed by the investors in our dataset as proxies for investor experience. These three experience-based variables are used as covariates, with the corporate transparency dummy variable, *Transpr_dv*, serving as the treatment variable. We employ a 1:1 nearest neighbor matching approach to match investors from the high and low transparency groups, based on similar investment experience, thereby mitigating the potential influence of investment experience on the analysis. The matched data are used in a grouped regression analysis, as shown in Table 7. The full sample consists of observations from both the high and low transparency groups. The disposition effect in the high transparency group is lower than in the low transparency group, and the difference in the regression coefficients for *Gain* between the two groups is significant at the 1% level. Thus, the results of the baseline regression are robust to concerns about sample selection bias.

Since the outbreak of the COVID-19 pandemic, a growing body of research has discussed the turbulence it caused in financial markets and its impact on investors (Schell et al., 2020, Al-Awadhi et al., 2020, Yuan et al., 2022, Ortmann et al., 2020, Xie et al., 2023). Studies show that during the pandemic, investors' psychology became more irrational and emotional, which directly affected their behavior(Xie et al., 2023). COVID-19, first identified in Wuhan, China, in December 2019, was attracting widespread attention in January 2020, and was declared a public health emergency of international concern. To test the robustness of our conclusions in light of the pandemic, we use 2020 as a cutoff point, treating data before 2020 as unaffected by the pandemic and data from 2020 onwards as impacted by the pandemic. We then perform grouped regressions for the "before" and "after" data, as shown in Table 8. The results show that in both groups, an increase in corporate transparency significantly reduces the disposition effect for investors, consistent with our baseline regression results. Investors exhibited a significantly greater disposition effect after the COVID-19 outbreak began, which aligns with our theoretical expectations, and the increase in transparency after the pandemic reduced the disposition effect even more compared to the pre-pandemic period.

The previous section considers all transactions and holding statuses after investors purchase stocks, including additional purchases, reductions in holdings (including complete liquidation), and the holding status as observations for the regression. Statistics show complete liquidations account for 65.96% of all sell transactions (including both reductions and liquidations), and 77.56% of the first sell transactions are complete liquidations. Here, we follow the approach in Lu et al. (2022) and consider only the period from the time an investor purchases a given stock until the first sell transaction involving that stock, to reflect how the profit or loss status of holding the stock affects individual investors' sell decisions. These results are shown in Table 9, the full sample includes stock transactions from companies with both high transparency and low levels of transparency. In the regression for the full sample, the coefficient for *Gain* is 0.3445, which is significantly greater than zero at the 1% level. This indicates that even when considering only the period from a stock purchase to the first sell transaction for that stock, the disposition effect remains significant. For the sample with high transparency companies only, the coefficient for *Gain* is 0.3266 and the average marginal effect is 0.0179. This indicates that when the investor's position in a stock of a high transparency company changes from a loss to a gain, the probability of selling the stock increases by 0.0179. In the sample of low transparency companies, the coefficient for *Gain* is 0.3951 and the average marginal effect is 0.0322. This suggests that when the investor's position in a stock of a low transparency company changes from a loss to a gain, the probability of selling the stock increases by 0.0322. The Suest test results indicate that the difference in the regression coefficients for *Gain* is significant at the 1% level, and that individual investors exhibit significant differences in terms of the disposition effect for stocks of companies with varying levels of transparency. Specifically, when the investor's position changes from a loss to a gain, stocks of companies with higher transparency have a 0.0143 lower probability of being sold compared to stocks of companies with lower transparency. In other words, individual investors exhibit a significantly smaller disposition effect for stocks of companies with higher transparency, and an increase in corporate transparency significantly reduces the disposition effect for individuals holding the company's stock, confirming our baseline result.

In the robustness test described above, we focus on the case where the investor purchases a stock and holds it until the first sale, which includes both position reductions and full liquidations. Here, we examine cases where investors fully liquidate a stock holding at the first sale. We repeat the same regression analysis and present the results in Table 10, which are consistent with the results in Table 9. In the full sample, the regression coefficient for *Gain* is 0.3377, which is significantly greater than 0 at the 1% level. This suggests the disposition effect remains significant even when considering only cases in which investors purchase a stock and fully liquidate the position at the first sale. In the high corporate transparency group, the average marginal effect of *Gain* is 0.0174, meaning that when the investor's position in a stock changes from a loss to a gain, the probability of selling increases by 0.0174. In the low corporate transparency group, the average marginal effect of *Gain* is 0.0312, indicating that when the investor's position in a stock changes from a loss to a gain, the probability of selling the stock increases by 0.0312. The average marginal effect of *Gain* in the high transparency group is 0.0138 lower than for the low transparency group, suggesting that when the investor's position changes from a loss to a gain, higher corporate transparency reduces the probability of selling the stock by 0.0138, compared to stocks of companies with lower transparency, reflecting a smaller disposition effect for stocks of companies with higher transparency. The results of the Suest test show the difference in the regression coefficients for *Gain* is statistically significant at the 1% level, suggesting the disposition effect for individual investors is significantly different for stocks of companies with varying levels of transparency. This confirms the robustness of our finding that an increase in corporate transparency significantly reduces the disposition effect.

### 4.4 Heterogeneity Analysis

Thus far, we have shown that corporate transparency significantly influences the disposition effect among individual investors. As transparency increases, the disposition effect exhibited by individual investors decreases significantly. Therefore, it is worth investigating whether a company's ownership structure impacts the effect of transparency on reducing the disposition effect.

We classify the stocks traded in our sample based on the ownership structure of their companies each year, categorizing them as state-owned enterprises (SOEs) and non-SOEs. Then, based on the median transparency of the companies in that year, we divide the trades into high and low transparency sub-groups. After grouping by ownership structure and transparency, we perform Probit regressions on the four groups. The results of these regressions and the average marginal effects for each group are presented in Table 11, along with inter-group difference tests for the regression coefficients, based on whether the enterprises are SOEs or non-SOEs. The results indicate that regardless of whether the traded stocks are issued by SOEs or non-SOEs, the average marginal effect of *Gain* is smaller in the high corporate transparency sub-group than in the low transparency sub-group. This suggests that individual investors exhibit a lower disposition effect for stocks of more transparent companies, regardless of ownership structure. Moreover, the difference in the disposition effect between stocks of companies with high versus low transparency levels is statistically significant at the 1% level. Finally, we compare the difference in the impact of corporate transparency on the disposition effect between the SOE and non-SOE groups. For SOEs, the regression coefficient of *Gain* in the high transparency sub-group decreased by 0.0808 and the average marginal effect of *Gain* decreased by 0.0157 relative to the low transparency sub-group. For non-SOEs, the regression coefficient of *Gain* in the high transparency sub-group decreased by 0.0698 and the average marginal effect of *Gain* decreased by 0.0141 compared to the low transparency sub-group. Therefore, we conclude that compared with the non-SOE group, an increase in corporate transparency in the SOE group has a greater impact on reducing the disposition effect for individual investors.

Next, we aggregate the total market capitalization at the end of the previous year for the companies whose stocks were traded each year in our sample, and use the median as the basis for classifying the companies into high and low market capitalization groups for that year. We then divide the trades into high and low transparency sub-groups, based on the median transparency of the companies in that year. After grouping by market capitalization and transparency, we perform Probit regressions on the four groups. The results of the regression and average marginal effects for each group are presented in Table 12, along with inter-group difference tests for the regression coefficients for different transparency sub-groups, based on whether the enterprises are in the high or low market capitalization groups. The results in Table 12 indicate that regardless of the market capitalization of the companies, the average marginal effect of *Gain* is smaller in the high transparency sub-group compared to the low transparency sub-group. This suggests that individual investors exhibit a lower disposition effect for stocks of companies with higher transparency, whether those companies are large or small. Moreover, the difference in the disposition effect between stocks of companies with varying transparency levels is statistically significant at the 1% level. Comparing the difference in the impact of corporate transparency on the disposition effect between the high-market-cap and low-market-cap groups, we find that in the high-market-cap group the coefficient for *Gain* decreased by 0.0764 and the average marginal effect of *Gain* decreased by 0.0125 relative to the low transparency sub-group. In the low-market-cap group, the coefficient for *Gain* decreased by 0.0327 and the average marginal effect of *Gain* decreased by 0.0051 relative to the low transparency sub-group. Compared to the low-market-cap group, the increase in corporate transparency in the high-market-cap group results in a more pronounced decline in the disposition effect for individual investors holding those stocks.

### 4.5 Mechanism Analysis

In the preceding sections we have consistently demonstrated that increased corporate transparency can reduce the disposition effect for individual investors; however, the underlying mechanisms for this finding remain unclear. To further explore how corporate transparency influences the disposition effect in individual stocks, we incorporate the corporate transparency indicator into our regression model and examine its impact on investor trading.

Research on the forms of the disposition effect (Lu et al., 2022, Ben-David and Hirshleifer, 2012) indicates that individual investors exhibit varying sensitivities when selling stocks with gains versus those with losses, which determines the magnitude of the disposition effect. Therefore, we hypothesize that the influence of corporate transparency on investor trading behavior may differ depending on the profit or loss status of the investors' positions. In this study, we examine the impact of corporate transparency on investor trading behavior under conditions of gains and losses, regardless of the magnitude of those gains or losses. We categorize investor trades into two groups based on the profit or loss status of the position prior to the transaction, the gain group and the loss group, and analyze how corporate transparency affects investor trading behavior in these two groups.

Here, we examine the impact of corporate transparency from two perspectives: the holding period and trading frequency when investors sell stocks. The holding period refers to the number of days (excluding non-trading days) from the time an investor opens a position in a particular stock until the position is fully liquidated (excluding partial sales). The trading frequency refers to the average number of trades per day (excluding non-trading days) involving that stock from the time an investor opens a position until it is fully liquidated, or until the end of the sample period if the position was not liquidated.

Table 13 presents the regression results regarding the impact of corporate transparency on investor selling behavior under different gain and loss conditions. We find that regardless of whether a stock is held at a gain or a loss, an increase in corporate transparency significantly reduces the probability that individual investors will sell the stock. Specifically, when a listed company exhibits higher transparency, the probability of individual investors selling stock positions that are profitable decreases, mitigating the disposition effect. Conversely, when a stock position is held at a loss, higher corporate transparency also reduces the probability of individual investors selling the stock, which exacerbates the disposition effect. By comparing the regression coefficients and average marginal effects of corporate transparency under both gain and loss conditions, it is evident that increased transparency has a more substantial effect in reducing the probability of selling stocks at a gain, thus overall diminishing the disposition effect.

It is worth noting that an increase in corporate transparency reduces the probability of selling both profitable and loss-making stocks. Correspondingly, it decreases and intensifies different components of the disposition effect, with a larger impact on the former and a smaller impact on the latter. A possible reason for this is that when a company is more transparent, the information it provides to the outside world becomes more comprehensive. This increased transparency helps investors gain a better understanding of the company's operational performance, thereby facilitating more informed and rational decision-making. Furthermore, based on their expectations regarding the company's future development, investors are more likely to adopt a long-term holding strategy, which reduces their sensitivity to short-term gains and losses. For companies with greater transparency, when the held stock is profitable, investors' confidence in holding it increases, and the probability of selling it decreases. When the held stock incurs a loss, investors' confidence in holding it weakens, but they often perceive the loss as temporary and maintain confidence in the company's long-term prospects, resulting in a reduced probability of selling the stock. It is precisely due to the asymmetric confidence shown towards profitable and losing stocks that the probability of selling profitable stocks is much lower than the probability of selling losing stocks, thereby reducing the disposition effect.

Table 14 presents regression results for the impact of corporate transparency on investors' holding period and trading frequency. Corporate transparency exerts a significant positive effect on the holding period, with a regression coefficient of 0.0886, which is statistically significant at the 1% level. This suggests that individual investors tend to hold stocks of more transparent companies for a significantly longer duration. Conversely, corporate transparency has a negative effect on trading frequency, with a coefficient of -0.0005, which is also significant at the 1% level. This indicates that individual investors tend to trade the stocks of more transparent companies significantly less frequently.

Based on the above analysis, we find that when a corporate transparency is high, regardless of whether the position held by individual investors is profitable or loss-making, those investors tend to hold the stock for a longer period rather than trading frequently. Specifically, long-term holding of profitable stocks reduces the disposition effect, while long-term holding of loss-making stocks exacerbates the disposition effect. We posit that greater corporate transparency helps investors gain a better understanding of the company's operational performance, thereby facilitating more informed and rational decision-making. Furthermore, based on their expectations regarding the company's future development, investors are more likely to adopt a long-term holding strategy, which reduces their sensitivity to short-term gains and losses. For companies with greater transparency, when the position in the stock is profitable, investors' confidence in holding it increases, reducing the bias toward selling profitable stocks. When the stock is held at a loss, investors' confidence in holding it weakens, but they often perceive the loss as temporary and maintain confidence in the company's long-term prospects, thus exacerbating the bias toward holding losing stocks. Comparing the effects of greater corporate transparency on the selling behavior involving profitable and losing stocks, the decline in the probability of selling profitable stocks is greater than the decline in the probability of selling losing stocks. The former reduces the disposition effect, while the latter increases it. Therefore, overall, an increase in corporate transparency significantly reduces the disposition effect.

## 5 Discussion

This study uses real trading portfolios on the Xueqiu platform to investigate the impact of corporate transparency on the disposition effect for individual investors, addressing a gap in the existing literature on this phenomenon. Previous research explores the existence (Odean, 1998, Locke and Mann, 2005, Cici, 2012, Wu et al., 2016, Wu et al., 2020), forms (Ben-David and Hirshleifer, 2012, Lu et al., 2022) of the disposition effect, and relationship between the disposition effect and investor characteristics (Wu et al., 2016, Xiao et al., 2018, Pereira da Silva and Mendes, 2021, Cao et al., 2021), and examines the influence of factors such as portfolio visibility (Danbolt et al., 2022), belief updating (Bachmann, 2024), and the profit and loss status of the portfolio (An et al., 2024) on the disposition effect. However, no studies explore the effect of corporate transparency on the disposition effect. Here, we first demonstrate the existence of the disposition effect for individual investors in China. Then, based on the quartiles of a corporate transparency index corresponding to the stocks traded by individual investors annually, we divide trades into four groups: low transparency, relatively low transparency, relatively high transparency, and high transparency. We conduct grouped regressions as the baseline analysis to examine the impact of corporate transparency on the disposition effect. The results indicate that higher corporate transparency reduces the disposition effect exhibited by individual investors, thereby supporting our research hypothesis. The reduction of the disposition effect contributes to the reversion of asset prices and enhances market efficiency. Therefore, the conclusion of our study aligns with the findings of previous research (Chen and Wu, 2022), both of which demonstrate that improvements in the information environment can enhance market efficiency. To verify the robustness of our baseline results, we conduct regressions that include interaction terms, group coefficient difference tests, and propensity score matching to address sample selection bias. We also consider the impact of the COVID-19 pandemic and conduct regressions using data based only on the first sale, and only those where all shares are sold in the first sale transaction. The results show the baseline results are robust. Heterogeneity tests indicate that improvement in corporate transparency has a greater effect in reducing the disposition effect for individual investors when stocks of SOEs and high market-cap companies are involved. Furthermore, a mechanism test reveals that for companies with greater transparency, when the investor's position in the stock is profitable, investors' confidence in holding the stock increases, leading to a reduced bias toward selling profitable stocks. When a stock is held at a loss, investors' confidence in holding the stock weakens, but they often perceive the loss as temporary and maintain confidence in the company's long-term prospects, thus exacerbating the bias toward holding losing stocks. Comparing the effects of increased corporate transparency on the selling behavior of profitable and losing stocks, the reduction in the probability of selling profitable stocks is greater than the reduction in the probability of selling losing stocks. The former reduces the disposition effect, while the latter exacerbates it. Therefore, overall, an increase in corporate transparency significantly reduces the disposition effect.

Previous literature has mainly examined the relationship between investor characteristics and the disposition effect (Wu et al., 2016, Xiao et al., 2018, Pereira da Silva and Mendes, 2021, Cao et al., 2021). In contrast, our study shifts the analytical focus from individual investor behavior to the corporate level. We demonstrate that corporate transparency is a key determinant influencing investor behavioral biases, thereby contributing to the existing body of research on the disposition effect. Furthermore, our findings offer practical insights for investors, firms, and regulatory bodies. For investors, our study suggests that, when confronted with companies of varying levels of transparency, opting for stocks of more transparent companies can mitigate the disposition effect. In the case of stock losses, it is essential for investors to promptly implement stop-loss strategies rather than relying on the belief that prices will likely rebound. Overcoming behavioral biases is crucial for making more rational investment decisions. For companies, enhancing transparency can reduce the abnormal stock price volatility stemming from irrational investor behavior. Moreover, increased transparency can bolster corporate reputation and market perception, thereby fostering long-term competitive advantages. Finally, for regulatory bodies, strengthening transparency regulations for publicly listed companies is recommended. This approach can mitigate irrational investor decision-making, thereby promoting the long-term stability of the market.

## 6 Conclusion

To our knowledge, this is the first study to examine the impact of corporate transparency on the disposition effect among individual investors. Our findings indicate that an increase in corporate transparency significantly reduces this disposition effect for a given stock. Furthermore, we examine the mechanisms through which increased transparency mitigates this effect. We find that for companies with greater transparency, when an investor's position in the stock is profitable, investors' confidence in holding the stock increases, leading to a reduced bias toward selling profitable stocks. When the stock is held at a loss, investors' confidence in holding the stock weakens, but they often perceive the loss as temporary and maintain confidence in the company's long-term prospects, thus exacerbating the bias toward holding losing stocks. The effect of increased transparency on the selling behavior of profitable stocks is greater than its effect on the selling behavior of losing stocks. Therefore, overall, an increase in corporate transparency significantly reduces the disposition effect.

This study contributes to a deeper understanding of the irrational behavior of individual investors. These investors make investment decisions based on past gains and losses, tending to sell profitable assets too early while holding onto losing assets for too long. This irrational behavior often makes it difficult to realize returns. Based pm the information environment of listed companies, this study offers new insights into this disposition effect, which is important for guiding individual investors to make more rational investment decisions and promoting the development of China's capital markets.

Of course, this study has its limitations. We use annual measures of corporate transparency, which cannot reflect the impact of changes in corporate transparency within a single year. If a higher-frequency method to measure corporate transparency could be found, the results might be more comprehensive. In addition, we use real trading portfolios on the Xueqiu platform, rather than account data from brokers. Xueqiu only provides the most recent 200 transactions for each real portfolio, so the sample data may not be comprehensive enough. If future research can overcome these limitations, the study will be richer and more convincing.

## Tables

**TABLE1** Variable Definitions.

| Variable Name | Variable Symbol | Meaning |
|---|---|---|
| Sell the Stock | *Sell* | When the transaction is a sale, it is defined as 1; otherwise, it is defined as 0. |
| Holding Profit | *Gain* | When the profit from holdings is positive, it is defined as 1; otherwise, it is defined as 0. |
| Earnings Quality | *EQ* | Absolute value of the discretionary accruals, calculated using the adjusted Jones model, multiplied by -1. |
| Information Disclosure Evaluation | *Score* | The Shanghai Stock Exchange (SSE) and Shenzhen Stock Exchange (SZSE) rate the information disclosure of listed companies, with the following scores: A for excellent (4 points), B for good (3 points), C for qualified (2 points), and D for unqualified (1 point). |
| Number of Analysts Covering the Company | *AF* | Number of analysts who forecasted the company's annual earnings that year. |
| Analysts' Earnings Forecasts Accuracy | *FA* | The median of the earnings per share forecasts from different analysts for the same year is first calculated, then the actual earnings per share is subtracted. This result is divided by the previous year's stock price per share, and the absolute value of this number is taken and multiplied by -1. |
| Audit Firm Quality | *Big4* | If the company's annual report auditor is one of the Big Four (PwC, KPMG, Deloitte, or EY), the value is 1; otherwise, it is 0. |
| Corporate transparency | *Transpr* | Average value of the percentiles corresponding to earnings quality, information disclosure evaluation, number of analysts covering the company, analysts' earnings forecast accuracy, and audit firm quality. |
| Corporate transparency dummy variable | *Transpr_dv* | The transparency is divided into high and low transparency based on the annual median of corporate transparency. When transparency is high, it is defined as 1; otherwise, it is defined as 0. |
| Number of Holding Days | *sqrt_TimeOwned* | Square root of the number of trading days of the stock from the first purchase to the current day. |
| Purchase Price | *log_BuyPrice* | Natural logarithm of the weighted average purchase price of the stock. |
| Stock Return Volatility | *Volatility* | Standard deviation of stock returns over the past 250 trading days. |
| Market Return | *MktRet* | Average logarithmic return of CSI 300 Index over past 20 trading days. |
| Market Return Volatility | *MktVol* | Standard deviation of CSI 300 Index logarithmic returns over past 20 trading days. |
| Company Size | *log_Size* | Natural logarithm of the total market capitalization at the end of the previous year. |
| Debt-to-Asset Ratio | *D/A ratio* | Debt-to-Asset Ratio at the end of the previous year. |
| Book-to-Market Ratio | *B/M ratio* | Market-to-Book Ratio at the end of the previous year. |
| Return on Assets | *ROA* | Return on Assets at the end of the previous year. |
| Ownership Structure | *OS* | Ownership structure of the current year. |
| Holding Period | *Hold_Period* | Number of trading days from the purchase date (day 0) to the sell date, excluding non-trading days. |
| Square Root of Holding Period | *sqrt_ Hold_Period* | Square root of the number of trading days from the purchase date (day 0) to the sell date, excluding non-trading days. |
| Trading Frequency | *Trade_Frequency* | Frequency of transactions made by the investor during the holding period of a stock, from the purchase to the sell date, excluding non-trading days. |

**TABLE 2** Summary statistics for the variables.

| Variable | Mean | St.dev | Min | 25th Pct | 50th Pct | 75th Pct | Max |
|---|---|---|---|---|---|---|---|
| *Sell* | 0.0337 | 0.1805 | 0 | 0 | 0 | 0 | 1 |
| *Gain* | 0.388 | 0.4873 | 0 | 0 | 0 | 1 | 1 |
| *Transpr* | 62.8567 | 17.4692 | 1.4248 | 50.748 | 64.1743 | 76.1313 | 96.8854 |
| *Transpr_dv* | 0.7847 | 0.4110 | 0.0000 | 1.0000 | 1.0000 | 1.0000 | 1.0000 |
| *sqrt_TimeOwned* | 11.2307 | 7.3416 | 1 | 5.2915 | 9.8995 | 15.906 | 40.5216 |
| *log_BuyPrice* | 2.9207 | 0.9057 | 0.1484 | 2.2086 | 2.8472 | 3.5107 | 7.7378 |
| *Volatility* | 0.0252 | 0.0094 | 0.0056 | 0.019 | 0.024 | 0.0301 | 0.1463 |
| *MktRet* | 0.0001 | 0.0026 | -0.0079 | -0.0016 | 0.0001 | 0.0016 | 0.0097 |
| *MktVol* | 0.0115 | 0.0044 | 0.0027 | 0.0084 | 0.0105 | 0.0141 | 0.0244 |
| *log_Size* | 17.7098 | 1.6632 | 12.676 | 16.3878 | 17.6614 | 19.0939 | 21.6692 |
| *D/A ratio* | 0.5521 | 0.7323 | 0.0084 | 0.35 | 0.5498 | 0.7493 | 178.3455 |
| *B/M ratio* | 0.6977 | 0.3058 | 0.0014 | 0.449 | 0.7281 | 0.9724 | 1.6433 |
| *ROA* | 0.0504 | 0.2291 | -30.6882 | 0.014 | 0.0387 | 0.0871 | 7.4461 |

**TABLE 3** Correlation analysis of variables.

| **Variable** | ***Sell*** | ***Gain*** | ***sqrt_Time Owned*** | ***log_Buy Price*** | ***Volatility*** | ***MktRet*** | ***MktVol*** | ***log_Size*** | ***D/A ratio*** | ***B/M ratio*** | ***ROA*** |
|---|---|---|---|---|---|---|---|---|---|---|---|
| *Sell* | 1 | | | | | | | | | | |
| *Gain* | 0.0632 | 1 | | | | | | | | | |
| *sqrt_TimeOwned* | -0.1598 | 0.0044 | 1 | | | | | | | | |
| *log_BuyPrice* | -0.0052 | 0.0661 | -0.0492 | 1 | | | | | | | |
| *Volatility* | 0.0864 | -0.0566 | -0.1890 | 0.1546 | 1 | | | | | | |
| *MktRet* | 0.0142 | 0.1437 | 0.0135 | -0.0213 | -0.0200 | 1 | | | | | |
| *MktVol* | -0.0239 | -0.0550 | 0.0964 | 0.0081 | 0.0145 | -0.2367 | 1 | | | | |
| *log_Size* | -0.0777 | 0.1740 | 0.1089 | 0.3576 | -0.3940 | -0.0266 | 0.0499 | 1 | | | |
| *D/A ratio* | -0.0113 | 0.0165 | 0.0299 | -0.0612 | -0.1034 | 0.0018 | 0.0075 | 0.1301 | 1 | | |
| *B/M ratio* | -0.0352 | -0.0168 | 0.1495 | -0.4620 | -0.4290 | 0.0317 | 0.0225 | 0.1586 | 0.1954 | 1 | |
| *ROA* | -0.0085 | 0.0266 | -0.0104 | 0.1390 | 0.0040 | -0.0063 | 0.0038 | 0.0735 | -0.5897 | -0.1192 | 1 |

**TABLE 4** Test of the individual investor disposition effect.

| **Variable** | ***Sell*** | | | |
|---|---|---|---|---|
| | **Probit** | **Probit_AME** | **Logit** | **Logit_AME** |
| *Gain* | 0.3564*** | 0.0226*** | 0.7527*** | 0.0220*** |
| | (0.0047) | (0.0003) | (0.01) | (0.0003) |
| *sqrt_TimeOwned* | -0.0767*** | -0.0049*** | -0.2059*** | -0.0060*** |
| | (0.0007) | (0.0001) | (0.0019) | (0.0001) |
| *log_BuyPrice* | 0.0302*** | 0.0019*** | 0.0668*** | 0.0020*** |
| | (0.0032) | (0.0002) | (0.0068) | (0.0002) |
| *Volatility* | 8.5764*** | 0.5427*** | 16.8245*** | 0.4923*** |
| | (0.2395) | (0.0154) | (0.4585) | (0.0135) |
| *MktRet* | 3.2168*** | 0.2036*** | 3.8685** | 0.1132** |
| | (0.7747) | (0.049) | (1.7245) | (0.0504) |
| *MktVol* | -1.2290** | -0.0778** | -3.0502** | -0.0893** |
| | (0.5862) | (0.0371) | (1.2709) | (0.0372) |
| *log_Size* | -0.0885*** | -0.0056*** | -0.1987*** | -0.0058*** |
| | (0.0019) | (0.0001) | (0.0042) | (0.0001) |
| *D/A ratio* | -0.0050*** | -0.0003*** | -0.0074*** | -0.0002*** |
| | (0.0017) | (0.0001) | (0.0029) | (0.0001) |
| *B/M ratio* | 0.0957*** | 0.0061*** | 0.1748*** | 0.0051*** |
| | (0.008) | (0.0005) | (0.017) | (0.0005) |
| *ROA* | -0.0327*** | -0.0021*** | -0.0476*** | -0.0014*** |
| | (0.0102) | (0.0006) | (0.0158) | (0.0005) |
| *Constant* | -0.2118*** | | 0.5178*** | |
| | (0.0359) | | (0.0759) | |
| *Observations* | 13730476 | 13730476 | 13730476 | 13730476 |
| *Pseudo R-squared* | 0.1450 | | 0.1554 | |

Note: The values in parentheses represent the robust standard errors clustered at the portfolio level. ***, **, and * represent p<0.01, p<0.05, and p<0.1, respectively. Probit_AME and Logit_AME denote the average marginal effects of the independent variables in the Probit and Logit models, respectively.

**TABLE 5** Test of the disposition effect on stocks with varying levels of corporate transparency.

| **Variable** | ***Sell*** | | | |
|---|---|---|---|---|
| | **Low transparency** | **Relatively low transparency** | **Relatively high transparency** | **High transparency** |
| *Gain* | 0.0373*** | 0.0334*** | 0.0259*** | 0.0170*** |
| | (0.0009) | (0.0008) | (0.0006) | (0.0003) |
| *sqrt_TimeOwned* | -0.0087*** | -0.0079*** | -0.0060*** | -0.0035*** |
| | (0.0002) | (0.0001) | (0.0001) | (0.0001) |
| *log_BuyPrice* | 0.0057*** | 0.0055*** | 0.0047*** | 0.0015*** |
| | (0.0007) | (0.0007) | (0.0004) | (0.0002) |
| *Volatility* | 0.5232*** | 0.5068*** | 0.5260*** | 0.4836*** |
| | (0.0381) | (0.0363) | (0.0273) | (0.0174) |
| *MktRet* | 0.4849*** | 0.3512*** | 0.0431 | 0.2330*** |
| | (0.1362) | (0.1212) | (0.0885) | (0.0446) |
| *MktVol* | 0.0712 | 0.059 | -0.0608 | -0.0936*** |
| | (0.0932) | (0.0829) | (0.0612) | (0.0324) |
| *log_Size* | -0.0084*** | -0.0077*** | -0.0068*** | -0.0031*** |
| | (0.0006) | (0.0005) | (0.0003) | (0.0001) |
| *D/A ratio* | 0.0001 | 0.0031 | -0.0057*** | -0.0080*** |
| | (0.0001) | (0.0022) | (0.0015) | (0.0009) |
| *B/M ratio* | -0.0005 | 0.0022 | 0.0043*** | 0.0019*** |
| | (0.0017) | (0.0018) | (0.0013) | (0.0007) |
| *ROA* | 0.0017** | -0.0208*** | -0.0557*** | -0.0683*** |
| | (0.0007) | (0.0064) | (0.0043) | (0.0032) |
| *Observations* | 1,384,772 | 1,490,691 | 2,485,580 | 8,361,595 |
| *Pseudo R-squared* | 0.1827 | 0.1685 | 0.1478 | 0.1169 |

Note: The table presents the average marginal effects of the independent variables in the Probit model, with the robust standard errors clustered at the portfolio level shown in parentheses. ***, **, and * represent p<0.01, p<0.05, and p<0.1, respectively.

**TABLE 6** Regression with an interaction term and a binary group.

| **Variable** | ***Sell*** | | | | | |
|---|---|---|---|---|---|---|
| | **Interaction term** | | **High transparency** | | **Low transparency** | |
| | ***Probit*** | ***Probit_AME*** | ***Probit*** | ***Probit_AME*** | ***Probit*** | ***Probit_AME*** |
| *Gain* | 0.4304*** | 0.0272*** | 0.3354*** | 0.0192*** | 0.4130*** | 0.0352*** |
| | (0.0074) | (0.0005) | (0.0048) | (0.0003) | (0.0075) | (0.0007) |
| *Transpr_dv* | -0.0138*** | -0.0009*** | | | | |
| | (0.0049) | (0.0003) | | | | |
| *Gain*Transpr_dv* | -0.0979*** | -0.0062*** | | | | |
| | (0.0069) | (0.0004) | | | | |
| *sqrt_TimeOwned* | -0.0765*** | -0.0048*** | -0.0700*** | -0.0040*** | -0.0970*** | -0.0083*** |
| | (0.0007) | (0.0001) | (0.0007) | (0.0001) | (0.0013) | (0.0001) |
| *log_BuyPrice* | 0.0349*** | 0.0022*** | 0.0364*** | 0.0021*** | 0.0559*** | 0.0048*** |
| | (0.0031) | (0.0002) | (0.0034) | (0.0002) | (0.0061) | (0.0005) |
| *Volatility* | 8.2595*** | 0.5220*** | 8.7844*** | 0.5017*** | 6.2434*** | 0.5320*** |
| | (0.2391) | (0.0153) | (0.2743) | (0.0158) | (0.3297) | (0.0284) |
| *MktRet* | 3.4410*** | 0.2175*** | 3.3634*** | 0.1921*** | 4.7792*** | 0.4073*** |
| | (0.7744) | (0.0489) | (0.7936) | (0.0453) | (1.2351) | (0.1051) |
| *MktVol* | -1.2431** | -0.0786** | -1.4178** | -0.0810** | 0.4348 | 0.0371 |
| | (0.586) | (0.037) | (0.589) | (0.0336) | (0.8657) | (0.0738) |
| *log_Size* | -0.0821*** | -0.0052*** | -0.0700*** | -0.0040*** | -0.0972*** | -0.0083*** |
| | (0.002) | (0.0001) | (0.0021) | (0.0001) | (0.0044) | (0.0004) |
| *D/A ratio* | -0.0048*** | -0.0003*** | -0.1112*** | -0.0063*** | 0.0002 | 0 |
| | (0.0017) | (0.0001) | (0.0137) | (0.0008) | (0.0014) | (0.0001) |
| *B/M ratio* | 0.0943*** | 0.0060*** | 0.0524*** | 0.0030*** | 0.0143 | 0.0012 |
| | (0.008) | (0.0005) | (0.0109) | (0.0006) | (0.0146) | (0.0012) |
| *ROA* | -0.0285*** | -0.0018*** | -1.0769*** | -0.0615*** | 0.007 | 0.0006 |
| | (0.01) | (0.0006) | (0.0465) | (0.0027) | (0.0076) | (0.0006) |
| *Constant* | -0.3166*** | | -0.4567*** | | 0.0906 | |
| | (0.0359) | | (0.0378) | | (0.077) | |
| *Observations* | 13,722,638 | 13,722,638 | 10,847,175 | 10,847,175 | 2,875,463 | 2,875,463 |
| *Pseudo R-squared* | 0.1454 | | 0.1284 | | 0.1754 | |
| *Suest test* | | | | chi2(1) = 122.27 | | |
| | | | | Prob > chi2 = 0.0000 | | |

Note: The table presents the average marginal effects of the independent variables in the Probit model, with the robust standard errors clustered at the portfolio level shown in parentheses. ***, **, and * represent p<0.01, p<0.05, and p<0.1, respectively.

**TABLE 7** Regression after propensity score matching.

| **Variable** | | | ***Sell*** | | | |
|---|---|---|---|---|---|---|
| | **Full sample** | | **High transparency** | | **Low transparency** | |
| | ***Probit*** | ***Probit_AME*** | ***Probit*** | ***Probit_AME*** | ***Probit*** | ***Probit_AME*** |
| *Gain* | 0.3564*** | 0.0225*** | 0.3352*** | 0.0191*** | 0.4133*** | 0.0351*** |
| | (0.0047) | (0.0003) | (0.0048) | (0.0003) | (0.0075) | (0.0007) |
| *sqrt_TimeOwned* | -0.0766*** | -0.0048*** | -0.0699*** | -0.0040*** | -0.0970*** | -0.0082*** |
| | (0.0007) | (0.0001) | (0.0007) | (0.0001) | (0.0013) | (0.0001) |
| *log_BuyPrice* | 0.0305*** | 0.0019*** | 0.0362*** | 0.0021*** | 0.0559*** | 0.0048*** |
| | (0.0032) | (0.0002) | (0.0034) | (0.0002) | (0.0061) | (0.0005) |
| *Volatility* | 8.5700*** | 0.5408*** | 8.7780*** | 0.5005*** | 6.2924*** | 0.5345*** |
| | (0.2413) | (0.0154) | (0.2758) | (0.0159) | (0.3316) | (0.0285) |
| *MktRet* | 3.2565*** | 0.2055*** | 3.4690*** | 0.1978*** | 4.5901*** | 0.3899*** |
| | (0.777) | (0.049) | (0.7966) | (0.0453) | (1.2334) | (0.1046) |
| *MktVol* | -1.2129** | -0.0765** | -1.4093** | -0.0803** | 0.4524 | 0.0384 |
| | (0.5874) | (0.0371) | (0.5905) | (0.0337) | (0.8622) | (0.0732) |
| *log_Size* | -0.0882*** | -0.0056*** | -0.0699*** | -0.0040*** | -0.0968*** | -0.0082*** |
| | (0.002) | (0.0001) | (0.0021) | (0.0001) | (0.0044) | (0.0004) |
| *D/A ratio* | -0.0047** | -0.0003** | -0.1113*** | -0.0063*** | 0.0004 | 0 |
| | (0.0022) | (0.0001) | (0.0137) | (0.0008) | (0.0017) | (0.0001) |
| *B/M ratio* | 0.0961*** | 0.0061*** | 0.0525*** | 0.0030*** | 0.0167 | 0.0014 |
| | (0.0081) | (0.0005) | (0.011) | (0.0006) | (0.0147) | (0.0012) |
| *ROA* | -0.0397*** | -0.0025*** | -1.0758*** | -0.0613*** | 0.0034 | 0.0003 |
| | (0.0118) | (0.0007) | (0.0467) | (0.0027) | (0.0086) | (0.0007) |
| *Constant* | -0.2201*** | | -0.4596*** | | 0.0777 | |
| | (0.0361) | | (0.038) | | (0.0773) | |
| *Observations* | 13,628,000 | 13,628,000 | 10,780,658 | 10,780,658 | 2,847,342 | 2,847,342 |
| *Pseudo R-squared* | 0.1445 | | 0.1280 | | 0.1752 | |
| *Suest test* | | | | chi2(1) = 122.48<br>Prob > chi2 = 0.0000 | | |

Note: The table presents the average marginal effects of the independent variables in the Probit model, with the robust standard errors clustered at the portfolio level shown in parentheses. ***, **, and * represent p<0.01, p<0.05, and p<0.1, respectively.

**Table 8** Regression before and after the Covid-19.

| **Variable** | ***Sell*** | | | | | | | |
|---|---|---|---|---|---|---|---|---|
| | **Before Covid-19** | | | | **After Covid-19** | | | |
| | **High transparency** | | **Low transparency** | | **High transparency** | | **Low transparency** | |
| | ***Probit*** | ***Probit_AME*** | ***Probit*** | ***Probit_AME*** | ***Probit*** | ***Probit_AME*** | ***Probit*** | ***Probit_AME*** |
| *Gain* | 0.2976*** | 0.0156*** | 0.3492*** | 0.0260*** | 0.3574*** | 0.0215*** | 0.4288*** | 0.0398*** |
| | 0.0063 | 0.0004 | 0.0104 | 0.0008 | 0.0063 | 0.0004 | 0.0098 | 0.001 |
| *sqrt_TimeOwned* | -0.0957*** | -0.0050*** | -0.1279*** | -0.0095*** | -0.0591*** | -0.0036*** | -0.0831*** | -0.0077*** |
| | 0.0012 | 0.0001 | 0.0022 | 0.0002 | 0.0008 | 0.0001 | 0.0015 | 0.0002 |
| *log_BuyPrice* | 0.0243*** | 0.0013*** | 0.0790*** | 0.0059*** | 0.0324*** | 0.0020*** | 0.0453*** | 0.0042*** |
| | 0.0048 | 0.0003 | 0.0096 | 0.0007 | 0.0044 | 0.0003 | 0.0079 | 0.0007 |
| *Volatility* | 6.6558*** | 0.3487*** | 5.4997*** | 0.4099*** | 9.2871*** | 0.5596*** | 5.8950*** | 0.5470*** |
| | 0.318 | 0.0167 | 0.378 | 0.0283 | 0.4417 | 0.0273 | 0.5763 | 0.054 |
| *MktRet* | 17.0527*** | 0.8933*** | 20.7908*** | 1.5497*** | -4.3158*** | -0.2601*** | -4.4786*** | -0.4156*** |
| | 1.3088 | 0.0681 | 2.0327 | 0.1511 | 1.011 | 0.0611 | 1.5332 | 0.1428 |
| *MktVol* | 6.7792*** | 0.3551*** | 10.7091*** | 0.7982*** | -7.9307*** | -0.4779*** | -9.3961*** | -0.8719*** |
| | 0.9272 | 0.049 | 1.3981 | 0.1055 | 0.7481 | 0.0454 | 1.1128 | 0.1042 |
| *log_Size* | -0.0736*** | -0.0039*** | -0.1060*** | -0.0079*** | -0.0753*** | -0.0045*** | -0.0839*** | -0.0078*** |
| | 0.0032 | 0.0002 | 0.007 | 0.0006 | 0.0028 | 0.0002 | 0.0052 | 0.0005 |
| *D/A ratio* | -0.0426** | -0.0022** | -0.0185 | -0.0014 | -0.0534*** | -0.0032*** | 0.0018 | 0.0002 |
| | 0.0214 | 0.0011 | 0.0231 | 0.0017 | 0.0171 | 0.001 | 0.0015 | 0.0001 |
| *B/M ratio* | -0.0772*** | -0.0040*** | -0.0459* | -0.0034* | 0.1066*** | 0.0064*** | 0.0315* | 0.0029* |
| | 0.0184 | 0.001 | 0.025 | 0.0019 | 0.0133 | 0.0008 | 0.0186 | 0.0017 |
| *ROA* | -1.2971*** | -0.0679*** | -0.0382 | -0.0028 | -0.7595*** | -0.0458*** | 0.0171** | 0.0016** |
| | 0.0888 | 0.0047 | 0.0259 | 0.0019 | 0.0532 | 0.0032 | 0.0084 | 0.0008 |
| *Constant* | -0.2023*** | | 0.2699** | | -0.4166*** | | -0.0333 | |
| | 0.0549 | | 0.1198 | | 0.0517 | | 0.0954 | |
| *Observations* | 4746136 | 4746136 | 1264698 | 1264698 | 6101039 | 6101039 | 1610765 | 1610765 |
| *Pseudo R-squared* | 0.1529 | | 0.2124 | | 0.1200 | | 0.1575 | |
| *Suest test* | | chi2(1) = 26.74<br>Prob > chi2 = 0 | | | | chi2(1) = 59.39<br>Prob > chi2 = 0 | | |

Note: The table presents the average marginal effects of the independent variables in the Probit model, with the robust standard errors clustered at the portfolio level shown in parentheses. ***, **, and * represent p<0.01, p<0.05, and p<0.1, respectively.

**TABLE 9** Only consider the first sale.

| **Variable** | ***Sell*** | | | | | |
|---|---|---|---|---|---|---|
| | **Full sample** | | **High transparency** | | **Low transparency** | |
| | **Probit** | **Probit_AME** | **Probit** | **Probit_AME** | **Probit** | **Probit_AME** |
| *Gain* | 0.3445*** | 0.0207*** | 0.3266*** | 0.0179*** | 0.3951*** | 0.0322*** |
| | (0.0047) | (0.0003) | (0.0048) | (0.0003) | (0.0077) | (0.0007) |
| *sqrt_TimeOwned* | -0.0764*** | -0.0046*** | -0.0694*** | -0.0038*** | -0.1019*** | -0.0083*** |
| | (0.0008) | (0.0001) | (0.0007) | (0.0001) | (0.0014) | (0.0001) |
| *log_BuyPrice* | 0.0243*** | 0.0015*** | 0.0295*** | 0.0016*** | 0.0522*** | 0.0043*** |
| | (0.0031) | (0.0002) | (0.0033) | (0.0002) | (0.0063) | (0.0005) |
| *Volatility* | 8.7747*** | 0.5279*** | 8.9110*** | 0.4884*** | 6.3688*** | 0.5196*** |
| | (0.2435) | (0.0149) | (0.2787) | (0.0155) | (0.337) | (0.0279) |
| *MktRet* | 4.7114*** | 0.2834*** | 4.5717*** | 0.2505*** | 6.8719*** | 0.5606*** |
| | (0.7787) | (0.0468) | (0.7947) | (0.0435) | (1.2949) | (0.1055) |
| *MktVol* | -0.346 | -0.0208 | -0.5617 | -0.0308 | 1.4806 | 0.1208 |
| | (0.5906) | (0.0355) | (0.5901) | (0.0323) | (0.9011) | (0.0735) |
| *log_Size* | -0.0755*** | -0.0045*** | -0.0578*** | -0.0032*** | -0.0820*** | -0.0067*** |
| | (0.002) | (0.0001) | (0.0021) | (0.0001) | (0.0044) | (0.0004) |
| *D/A ratio* | -0.0048*** | -0.0003*** | -0.1113*** | -0.0061*** | 0.0002 | 0 |
| | (0.0016) | (0.0001) | (0.0139) | (0.0008) | (0.0014) | (0.0001) |
| *B/M ratio* | 0.0963*** | 0.0058*** | 0.0458*** | 0.0025*** | 0.0242 | 0.002 |
| | (0.0081) | (0.0005) | (0.011) | (0.0006) | (0.015) | (0.0012) |
| *ROA* | -0.0325*** | -0.0020*** | -1.0618*** | -0.0582*** | 0.0051 | 0.0004 |
| | (0.0101) | (0.0006) | (0.047) | (0.0026) | (0.0077) | (0.0006) |
| *Constant* | -0.4540*** | | -0.6773*** | | -0.1682** | |
| | (0.0366) | | (0.0385) | | (0.0779) | |
| *Observations* | 12,824,106 | 12,824,106 | 10,388,947 | 10,388,947 | 2,429,196 | 2,429,196 |
| *Pseudo R-squared* | 0.1399 | | 0.1234 | | 0.1776 | |
| *Suest test* | | | | chi2(1) = 88.64<br>Prob > chi2 = 0.0000 | | |

Note: The values in the Probit column represent the regression coefficients of the Probit model. The Probit_AME column represents the average marginal effects of the independent variables in the Probit model. The values in parentheses are the robust standard errors clustered at the portfolio level. The Suest test is used to test the differences in regression coefficients between groups in the Probit model. ***, **, and * represent p<0.01, p<0.05, and p<0.1, respectively.

**TABLE 10** Only consider selling all at once on the first sale.

| **Variable** | | | ***Sell*** | | | |
|---|---|---|---|---|---|---|
| | **Full sample** | | **High transparency** | | **Low transparency** | |
| | **Probit** | **Probit_AME** | **Probit** | **Probit_AME** | **Probit** | **Probit_AME** |
| *Gain* | 0.3377*** | 0.0199*** | 0.3217*** | 0.0174*** | 0.3839*** | 0.0312*** |
| | (0.0047) | (0.0003) | (0.0049) | (0.0003) | (0.0078) | (0.0007) |
| *sqrt_TimeOwned* | -0.0759*** | -0.0045*** | -0.0690*** | -0.0037*** | -0.1032*** | -0.0084*** |
| | (0.0008) | (0.0001) | (0.0007) | (0.0001) | (0.0015) | (0.0001) |
| *log_BuyPrice* | 0.0252*** | 0.0015*** | 0.0293*** | 0.0016*** | 0.0548*** | 0.0044*** |
| | (0.0031) | (0.0002) | (0.0033) | (0.0002) | (0.0065) | (0.0005) |
| *Volatility* | 9.0025*** | 0.5318*** | 9.1159*** | 0.4917*** | 6.5293*** | 0.5306*** |
| | (0.2485) | (0.0149) | (0.2851) | (0.0156) | (0.3502) | (0.0289) |
| *MktRet* | 5.3902*** | 0.3184*** | 5.1864*** | 0.2797*** | 7.8057*** | 0.6343*** |
| | (0.7877) | (0.0465) | (0.8029) | (0.0432) | (1.3547) | (0.11) |
| *MktVol* | -0.1707 | -0.0101 | -0.4418 | -0.0238 | 1.8805** | 0.1528** |
| | (0.5962) | (0.0352) | (0.5947) | (0.0321) | (0.9371) | (0.0762) |
| *log_Size* | -0.0778*** | -0.0046*** | -0.0590*** | -0.0032*** | -0.0909*** | -0.0074*** |
| | (0.002) | (0.0001) | (0.0022) | (0.0001) | (0.0046) | (0.0004) |
| *D/A ratio* | -0.0048*** | -0.0003*** | -0.1114*** | -0.0060*** | 0.0001 | 0 |
| | (0.0016) | (0.0001) | (0.0143) | (0.0008) | (0.0013) | (0.0001) |
| *B/M ratio* | 0.0994*** | 0.0059*** | 0.0417*** | 0.0023*** | 0.0275* | 0.0022* |
| | (0.0082) | (0.0005) | (0.0111) | (0.0006) | (0.0157) | (0.0013) |
| *ROA* | -0.0321*** | -0.0019*** | -1.0849*** | -0.0585*** | 0.0059 | 0.0005 |
| | (0.01) | (0.0006) | (0.0482) | (0.0026) | (0.0077) | (0.0006) |
| *Constant* | -0.4232*** | | -0.6580*** | | -0.0233 | |
| | (0.038) | | (0.0399) | | (0.0813) | |
| *Observations* | 12,210,023 | 12,210,023 | 10,040,040 | 10,040,040 | 2,164,332 | 2,164,332 |
| *Pseudo R-squared* | 0.1397 | | 0.1227 | | 0.1817 | |
| *Suest test* | | | | | chi2(1) = 69.13<br>Prob > chi2 = 0.0000 | |

Note: The values in the Probit column represent the regression coefficients of the Probit model. The Probit_AME column represents the average marginal effects of the independent variables in the Probit model. The values in parentheses are the robust standard errors clustered at the portfolio level. The Suest test is used to test the differences in regression coefficients between groups in the Probit model. ***, **, and * represent p<0.01, p<0.05, and p<0.1, respectively.

**TABLE 11** The difference in the impact of transparency on the disposition effect of stocks in companies of different ownership structures.

| **Variable** | | | | | ***Sell*** | | | |
|---|---|---|---|---|---|---|---|---|
| | | **SOEs** | | | | **non-SOEs** | | |
| | **High transparency** | | **Low transparency** | | **High transparency** | | **Low transparency** | |
| | **Probit** | **Probit_AME** | **Probit** | **Probit_AME** | **Probit** | **Probit_AME** | **Probit** | **Probit_AME** |
| *Gain* | 0.3388*** | 0.0185*** | 0.4196*** | 0.0342*** | 0.3413*** | 0.0220*** | 0.4111*** | 0.0361*** |
| | (0.0058) | (0.0004) | (0.0105) | (0.0009) | (0.0059) | (0.0004) | (0.0086) | (0.0008) |
| *sqrt_TimeOwned* | -0.0694*** | -0.0038*** | -0.0935*** | -0.0076*** | -0.0748*** | -0.0048*** | -0.0999*** | -0.0088*** |
| | (0.0009) | (0.0001) | (0.002) | (0.0002) | (0.001) | (0.0001) | (0.0015) | (0.0002) |
| *log_BuyPrice* | 0.0512*** | 0.0028*** | 0.0757*** | 0.0062*** | 0.0385*** | 0.0025*** | 0.0591*** | 0.0052*** |
| | (0.0052) | (0.0003) | (0.0114) | (0.0009) | (0.0048) | (0.0003) | (0.0068) | (0.0006) |
| *Volatility* | 10.3770*** | 0.5661*** | 7.9401*** | 0.6478*** | 8.0585*** | 0.5189*** | 5.5391*** | 0.4860*** |
| | (0.4203) | (0.0231) | (0.6287) | (0.0519) | (0.335) | (0.022) | (0.3806) | (0.0336) |
| *MktRet* | 4.7704*** | 0.2602*** | 9.3762*** | 0.7650*** | 1.2412 | 0.0799 | 2.8944** | 0.2539** |
| | (0.9988) | (0.0544) | (1.7913) | (0.146) | (1.0055) | (0.0647) | (1.3948) | (0.1223) |
| *MktVol* | -1.9249*** | -0.1050*** | 4.7892*** | 0.3907*** | -1.1662 | -0.0751 | -1.5967* | -0.1401* |
| | (0.7188) | (0.0392) | (1.1911) | (0.0973) | (0.7108) | (0.0458) | (0.9659) | (0.0847) |
| *log_Size* | -0.0720*** | -0.0039*** | -0.0844*** | -0.0069*** | -0.0707*** | -0.0046*** | -0.1072*** | -0.0094*** |
| | (0.0028) | (0.0002) | (0.0067) | (0.0006) | (0.0031) | (0.0002) | (0.0052) | (0.0005) |
| *D/A ratio* | -0.1360*** | -0.0074*** | 0.0181 | 0.0015 | -0.1275*** | -0.0082*** | 0.0008 | 0.0001 |
| | (0.0193) | (0.0011) | (0.03) | (0.0024) | (0.0207) | (0.0013) | (0.0014) | (0.0001) |
| *B/M ratio* | 0.0590*** | 0.0032*** | -0.0099 | -0.0008 | 0.0670*** | 0.0043*** | 0.0129 | 0.0011 |
| | (0.0159) | (0.0009) | (0.0249) | (0.002) | (0.0157) | (0.001) | (0.0183) | (0.0016) |
| *ROA* | -1.4487*** | -0.0790*** | -0.2718*** | -0.0222*** | -0.9803*** | -0.0631*** | 0.0128* | 0.0011* |
| | (0.0798) | (0.0044) | (0.0745) | (0.0061) | (0.0548) | (0.0036) | (0.0076) | (0.0007) |
| *Constant* | -0.4731*** | | -0.2626** | | -0.4137*** | | 0.2939*** | |
| | (0.0498) | | (0.113) | | (0.0505) | | (0.0905) | |
| *Observations* | 4,832,127 | 4,832,127 | 951,724 | 951,724 | 4,689,457 | 4,689,457 | 1,794,203 | 1,794,203 |
| *Pseudo R-squared* | 0.1256 | | 0.1721 | | 0.1316 | | 0.1770 | |
| *Suest test* | | chi2(1) = 57.18<br>Prob > chi2 = 0.0000 | | | | chi2(1) = 67.01<br>Prob > chi2 = 0.0000 | | |

Note: The values in the Probit column represent the regression coefficients of the Probit model. The Probit_AME column represents the average marginal effects of the independent variables in the Probit model. The values in parentheses are the robust standard errors clustered at the portfolio level. The Suest test is used to test the differences in regression coefficients between groups in the Probit model. ***, **, and * represent p<0.01, p<0.05, and p<0.1, respectively.

**TABLE 12** The difference in the impact of transparency on the disposition effect of stocks in companies of different sizes.

| **Variable** | | | | | ***Sell*** | | | |
|---|---|---|---|---|---|---|---|---|
| | | **high market cap** | | | | **low market cap** | | |
| | **High transparency** | | **Low transparency** | | **High transparency** | | **Low transparency** | |
| | **Probit** | **Probit_AME** | **Probit** | **Probit_AME** | **Probit** | **Probit_AME** | **Probit** | **Probit_AME** |
| *Gain* | 0.3182*** | 0.0173*** | 0.3947*** | 0.0298*** | 0.4028*** | 0.0400*** | 0.4355*** | 0.0451*** |
| | (0.0048) | (0.0003) | (0.0082) | (0.0007) | (0.0116) | (0.0012) | (0.0107) | (0.0012) |
| *sqrt_TimeOwned* | -0.0678*** | -0.0037*** | -0.0886*** | -0.0067*** | -0.1083*** | -0.0107*** | -0.1119*** | -0.0116*** |
| | (0.0007) | (0.0001) | (0.0014) | (0.0001) | (0.0026) | (0.0003) | (0.0022) | (0.0003) |
| *log_BuyPrice* | -0.0052* | -0.0003* | 0.0230*** | 0.0017*** | 0.0979*** | 0.0097*** | 0.0739*** | 0.0076*** |
| | (0.0032) | (0.0002) | (0.007) | (0.0005) | (0.012) | (0.0012) | (0.0095) | (0.001) |
| *Volatility* | 11.1551*** | 0.6051*** | 6.6726*** | 0.5034*** | 7.8230*** | 0.7759*** | 5.6873*** | 0.5884*** |
| | (0.2845) | (0.0159) | (0.3885) | (0.0298) | (0.6368) | (0.0637) | (0.5193) | (0.0539) |
| *MktRet* | 5.2713*** | 0.2860*** | 10.3132*** | 0.7780*** | -3.3445 | -0.3317 | 0.7818 | 0.0809 |
| | (0.8095) | (0.0438) | (1.3869) | (0.1048) | (2.0605) | (0.2045) | (1.8183) | (0.1881) |
| *MktVol* | -1.9988*** | -0.1084*** | 3.0102*** | 0.2271*** | 0.0017 | 0.0002 | 1.2027 | 0.1244 |
| | (0.5964) | (0.0324) | (0.9559) | (0.0721) | (1.3693) | (0.1358) | (1.2527) | (0.1297) |
| *D/A ratio* | -0.2960*** | -0.0161*** | 0.0003 | 0 | -0.0995*** | -0.0099*** | -0.0013 | -0.0001 |
| | (0.0137) | (0.0008) | (0.0194) | (0.0015) | (0.0363) | (0.0036) | (0.0018) | (0.0002) |
| *B/M ratio* | 0.0375*** | 0.0020*** | -0.0286 | -0.0022 | 0.1709*** | 0.0169*** | 0.0388* | 0.0040* |
| | (0.0112) | (0.0006) | (0.0186) | (0.0014) | (0.0305) | (0.003) | (0.022) | (0.0023) |
| *ROA* | -1.2683*** | -0.0688*** | -0.0044 | -0.0003 | -1.1786*** | -0.1169*** | -0.005 | -0.0005 |
| | (0.0463) | (0.0025) | (0.0105) | (0.0008) | (0.1801) | (0.018) | (0.0115) | (0.0012) |
| *Constant* | -1.5372*** | | -1.5006*** | | -1.5195*** | | -1.3571*** | |
| | (0.0196) | | (0.0326) | | (0.048) | | (0.04) | |
| *Observations* | 10,174,017 | 10,174,017 | 1,879,129 | 1,879,129 | 673,158 | 673,158 | 996,334 | 996,334 |
| *Pseudo R-squared* | 0.1138 | | 0.1551 | | 0.1886 | | 0.1908 | |
| *Suest test* | | | chi2(1) = 92.29 | | | | chi2(1) = 6.36 | |
| | | | Prob > chi2 = 0.0000 | | | | Prob > chi2 = 0.0117 | |

Note: The values in the Probit column represent the regression coefficients of the Probit model. The Probit_AME column represents the average marginal effects of the independent variables in the Probit model. The values in parentheses are the robust standard errors clustered at the portfolio level. The Suest test is used to test the differences in regression coefficients between groups in the Probit model. ***, **, and * represent p<0.01, p<0.05, and p<0.1, respectively.

**TABLE 13** The impact of corporate transparency on the selling behavior of stocks with gains or losses.

| **Variable** | ***Sell*** | | | |
|---|---|---|---|---|
| | **Gain** | | **Loss** | |
| | **Probit** | **Probit_AME** | **Probit** | **Probit_AME** |
| *Transpr* | -0.0017*** | $-1.475\times10^{-4}$*** | -0.0013*** | $-6.34\times10^{-5}$*** |
| | (0.0002) | (0) | (0.0002) | (0) |
| *sqrt_TimeOwned* | -0.0731*** | -0.0062*** | -0.0802*** | -0.0039*** |
| | (0.0008) | (0.0001) | (0.0009) | (0.0001) |
| *log_BuyPrice* | 0.0233*** | 0.0020*** | 0.0500*** | 0.0024*** |
| | (0.004) | (0.0003) | (0.0036) | (0.0002) |
| *Volatility* | 10.7955*** | 0.9171*** | 5.6840*** | 0.2783*** |
| | (0.3187) | (0.0276) | (0.2599) | (0.0129) |
| *MktRet* | -2.4018*** | -0.2040*** | 10.3015*** | 0.5045*** |
| | (0.9) | (0.0766) | (0.964) | (0.0472) |
| *MktVol* | 0.6683 | 0.0568 | -2.8894*** | -0.1415*** |
| | (0.6642) | (0.0564) | (0.691) | (0.0338) |
| *log_Size* | -0.0856*** | -0.0073*** | -0.0729*** | -0.0036*** |
| | (0.0027) | (0.0002) | (0.0023) | (0.0001) |
| *D/A ratio* | -0.0050** | -0.0004** | -0.0027 | -0.0001 |
| | (0.0025) | (0.0002) | (0.002) | (0.0001) |
| *B/M ratio* | 0.1081*** | 0.0092*** | 0.0998*** | 0.0049*** |
| | (0.0106) | (0.0009) | (0.0091) | (0.0004) |
| *ROA* | -0.0270* | -0.0023* | -0.0305*** | -0.0015*** |
| | (0.0148) | (0.0013) | (0.0117) | (0.0006) |
| *Constant* | 0.1127** | | -0.3395*** | |
| | (0.0454) | | (0.0404) | |
| *Observations* | 5374298 | 5374298 | 8321617 | 8321617 |
| *Pseudo R-squared* | 0.1411 | | 0.1262 | |

Note: The values in the Probit column represent the regression coefficients of the Probit model. The Probit_AME column represents the average marginal effects of the independent variables in the Probit model. The values in parentheses are the robust standard errors clustered at the portfolio level. ***, **, and * represent p<0.01, p<0.05, and p<0.1, respectively.

**TABLE 14** The impact of transparency on holding period and trading frequency.

| Variable | *Hold_Period* | *sqrt_ Hold_Period* | *Trade_Frequency* |
|---|---|---|---|
| *Transpr* | 0.0886*** | 0.0053*** | -0.0005*** |
| | (0.0149) | (0.0007) | (0.0001) |
| *StkRet* | -191.6651*** | -14.7851*** | 2.1548*** |
| | (6.3721) | (0.422) | (0.0777) |
| *StkVol* | -327.0714*** | -21.6993*** | 3.1148*** |
| | (8.3605) | (0.4534) | (0.0677) |
| *MktRet* | -15.0521 | -2.3141 | -0.3621 |
| | (24.5649) | (1.6257) | (0.2763) |
| *MktVol* | 1987.4195*** | 115.6469*** | -12.0826*** |
| | (53.6161) | (2.909) | (0.3563) |
| *log_Size* | 8.9739*** | 0.5441*** | -0.0580*** |
| | (0.2756) | (0.0132) | (0.0015) |
| *D/A ratio* | 0.2799** | 0.0264*** | -0.0046** |
| | (0.1336) | (0.0092) | (0.002) |
| *B/M ratio* | 21.2285*** | 0.8513*** | -0.0351*** |
| | (1.1984) | (0.0558) | (0.0055) |
| *ROA* | 1.5208* | 0.1482*** | -0.0250*** |
| | (0.7903) | (0.0529) | (0.0075) |
| *Constant* | -139.5349*** | -6.4156*** | 1.6151*** |
| | (4.6102) | (0.2172) | (0.0249) |
| *Observations* | 301825 | 301825 | 310554 |
| *R-squared* | 0.0474 | 0.0880 | 0.1008 |

Note: The table shows the results of the OLS regression. The values in parentheses are the robust standard errors clustered at the portfolio level. ***, **, and * represent p<0.01, p<0.05, and p<0.1, respectively.

## Data Availability Statement

The raw data supporting the conclusions of this article will be made available by the authors, without undue reservation.

## Author Contributions

SC: Conceptualization, Data curation, Formal analysis, Investigation, Methodology, Software, Validation, Visualization, Writing – original draft, Writing – review & editing.

FR: Funding acquisition, Project administration, Resources, Supervision, Writing – review & editing.

## Funding

This research was funded by National Natural Science Foundation of China, grant number 71871094.

**Conflict of Interest**

The authors declare that the research was conducted in the absence of any commercial or financial relationships that could be construed as a potential conflict of interest.